A résumé of the work:

# Three Logistic Models for the Ecological and Economic Interactions: Symbiosis, Predator-Prey and Competition [*]

## Abstract

If one isolated species (corporation) is supposed to evolve following the logistic mapping, then we are tempted to think that the dynamics of two species (corporations) can be expressed by a coupled system of two discrete logistic equations. As three basic relationships between two species are present in Nature, namely symbiosis, predator-prey and competition, three different models are obtained. Each model is a cubic two-dimensional discrete logistic-type equation with its own dynamical properties: stationary regime, periodicity, quasi-periodicity and chaos. We also propose that these models could be useful for thinking in the different interactions happening in the economic world, as for instance for the competition and the collaboration between corporations. Furthermore, these models could be considered as the basic ingredients to construct more complex interactions in the ecological and economic networks.

# Introduction

Malthus claimed that human population $p$ growths faster than food supplies and then poverty would be the inevitable result of overpopulation (Malthus, 1798). He did not envisage the existence of a bound in the geometrical population growth given by the linear differential equation:

$$\frac{dp}{dt} = kp,$$

with $k$ the growth rate. Verhulst argued such limit as consequence of the available material: land and food. He established the simplest hypothesis, namely that the growth coefficient is proportional to the distance of the population size from its saturation point (Verhulst, 1845). The result is an inhibitory term $np^2$ proportional to the square of the population size:

$$\frac{dp}{dt} = kp - np^2,$$

which tends asymptotically to the constant population $k/n$. Verhulst called this equation the *logistic function*. Until today, nobody knows the reason for this name although an explanation with the military meaning of the word 'logistic' seems reasonable. We must wait until next century to see the importance of the Verhulst's work. It was the biologist Robert May (1976) who stated that the understanding of the discrete logistic model should be considered a milestone in the field of nonlinear phenomena. In concrete, the logistic equation is the basis of modern chaos theory, namely the paradigm for the period-doubling cascade (Feigenbaum, 1978) and also the paradigm of a particular type of intermittence (Pomeau & Manneville, 1980).

Let us state the discrete logistic equation, that is, the discrete version of Verhulst model, for the evolution of the population of a species (in ecology) or corporation (in economics). In the following we restrict ourselves to the ecological interpretation, although all this work would be equivalent in the economic world where corporations interact as species do in an ecosystem. Thus, if $x_n$ represents the population of an isolated species after $n$ generations, let us suppose this variable is bounded in the range $0 < x_n < 1$. The discrete logistic evolution is given by the equation

$$x_{n+1} = \mu x_n (1 - x_n), \qquad (1)$$

where $0 < \mu < 4$ in order to assure $0 < x_n < 1$. The *activation or expanding phase* is controlled by the term $\mu x_n$ proportional to the current population $x_n$ and to the constant *growth rate* $\mu$. Resource limitations bring the system to an *inhibition or contracting phase* directly related with overpopulation. The term $(1 - x_n)$ can denote how far the system is of overcrowding. Therefore, if we take the product of both terms as the simplest approach to the population dynamics, equation (1) results to be the equation modelling the rich dynamics of an isolated species with finite affordable resources. The dynamical behaviour that is found when the growth rate is modified is as follows (Collet & Eckmann, 1980 ; Mira, 1987):

(i)     $0 < \mu < 1$: The growth rate is not big enough to stabilize the population. It will drop and the specie will become extinct.

(ii) $1 < \mu < 3$: A drastic change is obtained when $\mu$ is greater than 1. The non vanishing equilibrium between the two competing forces, reproduction on one hand and resource limitation on the other, is now possible. The population reaches, independently of its initial conditions, a fixed value that is maintained in time.

(iii) $3 < \mu < 3.57$: A cascade of sudden changes provokes that the population oscillates in cycles of period $2^n$, where $n$ increases from 1, when $\mu$ is close to 3, to infinity when $\mu$ is approaching the critical value 3.57. This is named the period-doubling cascade.

(iv) $3.57 < \mu < 3.82$: When the parameter moves, the system alternates between periodical behaviours with high periods on parameter interval windows and *chaotic regimes* for parameter values not located in intervals. The population can be not predictable although the system is deterministic. The chaotic regimes are observed for a given value of μ on sub-intervals of [0,1].

(v) $3.82 < \mu < 3.85$: The orbit of period 3 appears for $\mu = 3.82$ after a regime where unpredictable bursts, named *intermittences*, have become rarer until their disappearance in the three-periodic time signal. The existence of the period 3 orbit means, such as the Sarkovskii theorem tells us, that all periods are possible for the population dynamics, although, in this case, they are not observable due to their instability. What it is observed in this range is the period-doubling cascade $3*2^n$.

(vi) $3.85 < \mu < 4$: Chaotic behaviour with periodic windows is observed in this interval.

(vii) $\mu = 4$: The chaotic regime is obtained on the whole interval [0,1]. This specific regime produces dynamics, which looks like random. The dynamics has lost near-all its determinism and the population evolves as a random number generator.

Therefore, there are essentially three remarkable dynamical behaviours in this system: the period doubling route to chaos around the value $\mu \approx 3.57$ (Feigenbaum, 1978), the time signal complexification by intermittence in the neighbourhood of $\mu \approx 3.82$ (Pomeau & Manneville, 1980) and the random-like dynamics for $\mu = 4$.

## **Background**

Let us suppose now, under a similar scheme of expansion/contraction, that two species $(x_n, y_n)$ are now living together. Each one of them evolves following a logistic-type dynamics,

$$x_{n+1} = \mu_x(y_n) x_n (1 - x_n),$$
$$y_{n+1} = \mu_y(x_n) y_n (1 - y_n). \quad (2)$$

The interaction between species causes the growth rate $\mu(z)$ to vary with time, then $\mu(z)$ depends on the population size of the others and on a positive constant $\lambda$, that measures the strength of the mutual interaction. The simplest choice for this growth rate can be a linear increasing $\mu_1$ or decreasing $\mu_2$ function expanding at the parameter interval where the logistic map shows some activity, that is, $\mu \in (1,4)$. Thus, we have

$$\mu_1(z) = \lambda(3z + 1),$$
$$\mu_2(z) = \lambda(-3z + 4).$$

Depending on the combination of both functions $\mu_1$ and $\mu_2$ we obtain three different models:

(1) The *symbiosis* between species can be modelled by the symmetrical coupling meaning a mutual interacting benefit, then $\mu_x = \mu_y = \mu_1$ (López-Ruiz & Fournier-Prunaret, 2004).
(2) The *predator-prey interaction* is based on the benefit/damage relationship established between the predator and prey, respectively, then $\mu_x = \mu_1$ and $\mu_y = \mu_2$ (López-Ruiz & Fournier-Prunaret, 2005).
(3) The *competition* between species causes the contrary symmetrical coupling, then $\mu_x = \mu_y = \mu_2$ (López-Ruiz & Fournier-Prunaret, 2006).

These three discrete two-dimensional logistic systems could also be considered for modelling the interaction in social or economical systems, changing in an equivalent form the species by agents or corporations, respectively. Moreover these models can be used in future studies as the bricks necessary to build more complex networks where different ecological, biological or economic interactions among *n*-species, *n*-agents or *n*-corporations are taking place.

# **Main Focus**

Let us explain in more detail the different dynamical regimes that can be found in the three models of the former section. Depending on the coupling parameter, steady states, periodic orbits or chaotic oscillations are displayed by the three models. It means that two coexisting species in ecology (Murray, 2002) or, equivalently, two interacting corporations (duopoly) in the market place (Cournot, 1838) show a rich variety of behaviours: mutual extinction, stationary coexistence, periodic cycles and chaotic unpredictability. The knowledge of the set of initial conditions (basin) leading to each one of those dynamical behaviours is also an important question (Mira et al., 1996). The basins for the different attractors present in the three models are depicted in the figures which are given through this article.

## **Symbiosis**

When two species $(x_n, y_n)$ interact symbiotically the equation (2) takes the form:

$$
\begin{aligned}
x_{n+1} &= \lambda(3y_n + 1)x_n(1 - x_n), \\
y_{n+1} &= \lambda(3x_n + 1)y_n(1 - y_n),
\end{aligned} \quad (3)
$$

where the positive constant $\lambda$ is called *the mutual benefit*. The numerical simulations show that the real and adjustable parameter $\lambda$ has sense in the range $0 < \lambda < 1.084$. This application can be represented by $T_\lambda : [0,1] \times [0,1] \to \Re^2$, $T_\lambda(x_n, y_n) = (x_{n+1}, y_{n+1})$. Let us observe that when $y_n = 0$ or $x_n = 0$, the logistic dynamics for one isolated specie is recovered. In this case the parameter $\lambda$ takes the role of the parameter $\mu$ of equation (1). When $\lambda$ is modified, the coupled logistic system (3) shows the following dynamical behaviour:

(i) $0 < \lambda < 0.75$: The mutual benefit is not big enough to allow a stable coexistence of both species and they will disappear toward $p_0 = (0,0)$. The whole square $[0,1] \times [0,1]$ is asymptotically shrunk to $p_0$. The total basin (red color) of $p_0$ is fractal (Fig. 1).

(ii) $0.75 < \lambda < 0.866$: A sudden change is obtained when $\lambda$ is greater than 0.75. Both populations are synchronized to a stable non-vanishing fixed quantity $p_4$ when the initial populations overcome certain critical values (yellow color). If the initial species are under these limits both will become extinct toward $p_0 = (0,0)$ (Fig. 2).

(iii) $0.866 < \lambda < 0.957$: The system is now bi-stable. Each one of the species oscillates out-of-phase between the same two fixed values $p_{5,6}$. This is a lag-synchronized state or in other words a stable 2-period orbit. There is still in this range the possibility of extinction toward $p_0 = (0,0)$ when the initial populations are very small or close to the overcrowding (Fig. 3).

(iv) $0.95 < \lambda < 1.03$: The system is not anymore on a periodic orbit. It acquires a new frequency and the dynamics is now quasiperiodic. Both populations oscillate among slightly infinitely many different states (a pair of invariant closed curves). Synchronization is lost. There are in this regime periodic windows where the system becomes newly lag-synchronized. Also, for initial populations nearly zero and for $\lambda < 1$, the species can not survive (Fig. 4).

(v) $1.03 < \lambda < 1.0843$: The system is now in a chaotic regime (Fig. 5a). It is characterized by a noisy-like small oscillation around a synchronized state with non-periodic unpredictable bursts. Periodic oscillations can also be obtained for some particular values of the mutual benefit. Some other initial conditions are not meaningful or interpretable in this scheme because the system is going outward the square $[0,1] \times [0,1]$ and evolves toward infinity. The system 'crashes'. This sudden 'damage' is interpreted as some kind of catastrophe provoking the extinction of species (Fig. 5b).

Let us remark that although the equations are formed by logistic-type components, the logistic effects have been lost and a completely new scenario emerges when they are coupled. In this case, the symbiotic interaction makes the species to reach different stable states. Depending on the mutual benefit, the system can reach the extinction, a fixed synchronized state, a bi-stable lag-synchronized configuration, an oscillating dynamics among infinitely many possible states or a chaotic regime. We must highlight in this model the transition to chaos by the Ruelle-Takens route (Eckmann, 1981). All these behaviours are caused by the symbiotic coupling of the species and are not predictable from the properties of the individual logistic evolution of each one of them. Moreover, this interaction implies a mutual profit for both species. In fact, when $\mu < 1$ one of the isolated species is extinct, but it can survive for $\lambda < 1$ if a small quantity of the other species is aggregated to the ecosystem. Hence, the symbiosis appears to be well held in this cubic model. We summarise the dynamical behaviour of model (3). The different parameter regions where the mapping $T$ has stable attractors are given in the next table.

| INTERVAL | NUMBER OF ATTRACTORS | ATTRACTORS |
|---|---|---|
| $0 < \lambda < 0.75$ | 1 | $p_0$ |
| $0.75 < \lambda < 0.866$ | 2 | $p_0$, $p_4$ |
| $0.866 < \lambda < 0.957$ | 2 | $p_0$, $p_{5,6}$ |
| $0.957 < \lambda < 1$ | 2 | $p_0$, pair of invariant closed curves |
| $1 < \lambda < 1.03$ | 1 | pair of invariant closed curves |

| 1.03 < λ < 1.0843 | 1 | symmetric chaotic attractor (or frequency lockings) |

## Predator-Prey

Let us think now that one of the species $x_n$, the predator, is supposed to benefit by attacking the other species $y_n$, the prey, and on the contrary, this last species $y_n$ suffers damage in its population as a consequence of the attacks of the other species $x_n$. Then both species are governed by an *attack-defence* interaction which takes place between them. It makes the parameter $\lambda$ to lose its pure reproductive meaning that it had in the case of isolation. It can now represent some kind of *mixed reproduction rate* containing also information on the mutual interaction between the species. This simplification facilitates our research in the sense that it is easier to explore with some detail the behaviour of the *final predator-prey model*,

$$x_{n+1} = \lambda(3y_n + 1)x_n(1 - x_n),$$
$$y_{n+1} = \lambda(-3x_n + 4)y_n(1 - y_n), \qquad (4)$$

as a function of the only real positive parameter $\lambda$, which has sense in the range $0 < \lambda < 1.21$. This application can also be represented by $T_\lambda : [0,1] \times [0,1] \to \Re^2$, $T_\lambda(x_n, y_n) = (x_{n+1}, y_{n+1})$. When $\lambda$ is modified, the coupled logistic system (4) of Lotka-Volterra type (Lotka, 1925 ; Volterra, 1926) behaves as follows:

(vi) $0 < \lambda < 0.25$: The reproductive force of the preys is smaller than the combination of its natural death rate and the effect of predator attacks. Hence preys can not survive and the predators become also extinct.

(vii) $0.25 < \lambda < 0.4375$: Prey population can survive in a small quantity but predators do not have enough food to be self-sustained and become extinct. This phenomenon is some kind of *indirect Allee effect* for the case of two species, that is, if the prey population does not reach a certain threshold the predator population decreases in size until the extinction.

(viii) $0.4375 < \lambda < 1.051$: When the prey population exceeds the threshold $y^* \approx 0.43$, predator's attack strategy is efficient and the system settles down in an equilibrium $p_4$ where both populations are fixed in time (Fig. 6). The increasing of $\lambda$ allows the coexistence of bigger populations: predator population increases faster than prey population up to reach a slightly greater density than that of the preys. When $p_4$ is around $x^* \approx y^* \approx 0.6$ for $\lambda = 1.051$ a new instability takes place.

(ix) $1.051 < \lambda < 1.0851$: For $\lambda = 1.051$, a stable period three cycle ($Q_1$, $Q_2$, $Q_3$) appears in the system (Fig. 7). It coexists with the former fixed point $p_4$. When $\lambda$ is increased, a period-doubling cascade takes place and generates successive cycles of higher periods $3 \cdot 2^n$. The system presents bistability. Depending on the initial conditions (red zone or yellow one), both populations oscillate in a periodic orbit or, alternatively, settle down in the fixed point.

(x) $1.0851 < \lambda < 1.0997$: In this region, an aperiodic dynamics is possible. The period-doubling cascade has finally given birth to an order three cyclic chaotic bands ($A_{31}$, $A_{32}$, $A_{33}$). The system can now present an irregular oscillation besides the stable equilibrium $p_4$ with final fixed populations (Fig. 8).

(xi) $1.0997 < \lambda < 1.1758$: The basin of attraction of chaotic bands is absorbed by the one of the fixed point $p_4$ and the populations reach always the constant equilibrium, although in some cases, after a chaotic transient (Fig. 9).

(xii) $1.1758 < \lambda < 1.211$: The system suffers a Hopf bifurcation giving rise to a stable invariant curve. The populations oscillate among a continuum of possible states located on the invariant curve (Fig. 10).

(xiii) $\lambda > 1.211$: The iterations are going outwards the square $[0,1] \times [0,1]$ and evolve towards infinity. The system 'crashes'. This critical value can be interpreted as some kind of catastrophe provoking the extinction of species.

Let us remark that there is no possible survival of predators without preys for a low reproduction rate. Only when the prey population reaches a threshold the predators can find an adequate hunting strategy to survive. This fact presents some similarity with the *Allee effect* for one species, that is, the decreasing in size of one species if this falls below a critical level. Another remarkable fact is the bistability between the fixed point and the high periodic orbits growing from the period doubling cascade of the period three cycle. And, finally, the possibility of a chaotic dynamics in the evolution of both species introduces an additional value in benefit of this model. Summarising, the different parameter regions where the mapping $T$ has stable attractors are given in the next table.

| INTERVAL | NUMBER OF ATTRACTORS | ATTRACTORS |
|---|---|---|
| $0 < \lambda < 0.25$ | 1 | $p_0$ |
| $0.25 < \lambda < 0.4375$ | 1 | $p_2$ |
| $0.4375 < \lambda < 1.051$ | 1 | $p_4$ |
| $1.051 < \lambda < 1.075$ | 2 | $p_4$, period three cycle |
| $1.075 < \lambda < 1.08511$ | 2 | $p_4$, period orbit of the doubling cascade $3 \cdot 2^n$ |
| $1.08511 < \lambda < 1.09967$ | 2 | $p_4$, period three chaotic bands |
| $1.09967 < \lambda < 1.17579$ | 1 | $p_4$ |
| $1.17579 < \lambda < 1.21091$ | 1 | invariant closed curve (also frequency locking) |

## Competition

Let us suppose now two species $(x_n, y_n)$ evolving under a competitive interaction (Darwin, 1859). The model for the competition is:

$$x_{n+1} = \lambda(-3y_n + 4)x_n(1 - x_n),$$
$$y_{n+1} = \lambda(-3x_n + 4)y_n(1 - y_n), \quad (5)$$

where the positive parameter $\lambda$ expresses the strength of the *mutual competitive interaction*, and has sense in the range $0 < \lambda < 1.21$. When some of the species is null the typical logistic behaviour is found for the other species. The system (5) can be represented by the application $T_\lambda : [0,1] \times [0,1] \to \Re^2$, $T_\lambda(x_n, y_n) = (x_{n+1}, y_{n+1})$, which represents a symmetric competition between two species that have access to the same amount of resources and that present a similar performance to compete for them. We explain now the dynamical behaviour of this coupled logistic system when $\lambda$ is modified. We obtain:

(xiv) $0 < \lambda < 0.25$: Although the competition is weak, the smallness of the growth rate provokes the extinction of species. We can also interpret that competition is an important force for survival in a pure competitive interaction and its absence or its weakness can cause extinction toward $p_0$.

(xv) $0.25 < \lambda < 0.9811$: Both populations can survive and the system settles down in a stationary symmetrical equilibrium $p_4$. This point is reached independently of the initial conditions. A greater $\lambda$ allows the coexistence of bigger populations.

(xvi) $0.9811 < \lambda < 1.1743$: For $\lambda = 0.9811$ the destabilization of the fixed point gives rise to a stable period two cycle located on the diagonal $(p_5, p_6)$. The populations oscillate now synchronously between the two values of that orbit.

(xvii) $1.1743 < \lambda < 1.1875$: The period two cycle becomes unstable and an off-diagonal period four orbit appears $(p_{51}, p_{52}, p_{61}, p_{62})$. What could seem the starting point of a period-doubling cascade is just its final point. No new flip bifurcation is found for this particular orbit.

(xviii) $1.1875 < \lambda < 1.1924$: A Neimark-Hopf bifurcation happens and the period four orbit gives rise to a period four invariant closed curve. The pattern of the time behaviour of the populations is now quasi-periodic.

(xix) $1.1924 < \lambda < 1.201$: The dynamics becomes complex, then chaotic, and a huge number of bifurcations happen in this regime. A detail is given in Fig. 11. In a global view, the populations visit successively in time an attractor which is formed by four chaotic rings (Fig. 12).

(xx) $1.201 < \lambda < 1.206$: A crisis of the four chaotic bands with the diagonal changes its aspect, which is now formed by two big chaotic bands (Fig. 13-14).

(xxi) $\lambda > 1.206$: The iterations are going outwards the square $[0,1] \times [0,1]$ and evolve towards infinity. This critical value can be interpreted as some kind of catastrophe provoking the extinction of species.

Let us remark that competition is a constructive force in this model. A weak competition can lead the system to extinction. A stronger interaction can stabilize a fixed population or can also generate periodic or chaotic oscillations. Apparently there is no multi-stability in the coarse unfolding above presented but, as we mention in (xix) a very localized zone in the parameter space, $1.1924 < \lambda < 1.1948$, is source of a fascinating dynamical richness; coexistence of high period orbits and their period doubling cascades succeed by slight modifications of the parameter $\lambda$. It reflects, if the biological simile is accepted, some kind of *explosion of life*. In a more general context of ecological networks, this fact puts in evidence that competition among species can enhance biodiversity, if this is interpreted as the different possibilities that Nature offers for the survival of species. For the sake of clarity, the different parameter regions where the mapping $T$ has stable attractors in the interval $0 < \lambda < 1.206$ are explicitly put in the next table.

| INTERVAL | NUMBER OF ATTRACTORS | ATTRACTORS |
| --- | --- | --- |
| $0 < \lambda < 0.25$ | 1 | $p_0$ |
| $0.25 < \lambda < 0.9811$ | 1 | $p_4$ |
| $0.9811 < \lambda < 1.1743$ | 1 | $(p_5, p_6)$ |
| $1.1743 < \lambda < 1.1875$ | 1 | $(p_{51}, p_{52}, p_{61}, p_{62})$ |
| $1.1875 < \lambda < 1.1924$ | 1 | period four invariant closed curve |
| $1.1924 < \lambda < 1.1948$ | multistability | high period orbits |
| $1.1948 < \lambda < 1.201$ | 1 | period four chaotic band |
| $1.201 < \lambda < 1.206$ | 1 | period two chaotic band |

## **Future Trends**

Ecological or economic interactions have fitness consequences for species or corporations, respectively. The possibility of new simple views, although unrealistic, on the basic interactions between species in an ecosystem or corporations in a particular market can enlarge the spectrum of combinations for building up more complex images of those ecosystems and markets. Following this path and using some of these new ingredients, different models with a large number of interacting elements will appear in the literature and will be one of the future trends in the field of complex systems. It is sure that these models can contribute and help to understand and to increase our insight about the evolution of ecological and corporation networks.

## **Conclusions**

Life in an ecosystem is a complex system. Depending on the scale of observation one can try to understand the interaction among individuals of the same species or one can dissert about the relationship among different species in a higher scale. In the first case we are doing sociology of a particular species and in the second case we are studying ecology. In any case, the detailed analysis of each problem reveals new variables outside the scale of observation that are important to explain the dynamical behaviour of the system. Thus, for instance, the human society can not be understood as only a set of interacting human beings without the constraints imposed by the social super-structures such as institutions, public administration, corporations, etc., and by a similar argument, the evolution of rabbits and foxes populations is not understandable independently of an entangled system of interactions with other species in their respective environments. Hence, as a first approach to the dynamics of an ecosystem, three main types of interaction between two species, namely the predator-prey situation, competition or mutualism, have been proposed as basic bricks for understanding how the populations evolve in that ecosystem. Thus, we have interpreted three cubic two-dimensional coupled logistic equations as three discrete models to explain either the evolution of two symbiotically interacting species, either a predator-prey system or two species under competitive interaction. Extinction, stable coexistence, periodic or chaotic oscillations are found in the dynamics of the three systems. Also the common phenomenon of multistability is present in the three models. We hope that the basic couplings here proposed can be useful for future investigations on multi-agent systems in order to better unveil the problem of human life in society.

## **Glossary**

ATTRACTING: An invariant subset of the plane is attracting if it has a neighbourhood every point of which tends asymptotically to that subset or arrives there in a finite number of iterations.

CHAOTIC AREA: An invariant subset that exhibits chaotic dynamics. A typical trajectory fills this area densely.

CHAOTIC ATTRACTOR: A chaotic area, which is attracting.

BASIN: The basin of attraction of an attracting set is the set of all points, which converge towards the attracting set.

IMMEDIATE BASIN: The largest connected part of a basin containing the attracting set.

CONTACT BIFURCATION: Bifurcation involving the contact between the boundaries of different regions. For instance, the contact between the boundary of a chaotic attractor and the boundary of its basin of attraction or the contact between a basin boundary and a critical curve *LC* are examples of this kind of bifurcation.

## **Figure Captions**

**Fig. 1 :** Model (3). See item (i) of the text for explanation.
**Fig. 2 :** Model (3). See item (ii) of the text for explanation.
**Fig. 3 :** Model (3). See item (iii) of the text for explanation.
**Fig. 4 :** Model (3). See item (iv) of the text for explanation.
**Fig. 5 :** Model (3). See item (v) of the text for explanation.
**Fig. 6 :** Model (4). See item (viii) of the text for explanation.
**Fig. 7 :** Model (4). See item (ix) of the text for explanation.
**Fig. 8 :** Model (4). See item (x) of the text for explanation.
**Fig. 9 :** Model (4). See item (xi) of the text for explanation.
**Fig. 10 :** Model (4). See item (xii) of the text for explanation.
**Fig. 11 :** Model (5). See item (xix) of the text for explanation.
**Fig. 12 :** Model (5). See item (xix) of the text for explanation.
**Fig. 13 :** Model (5). See item (xx) of the text for explanation.
**Fig. 14 :** Model (5). See item (xx) of the text for explanation.

\*\*\*\*\*\*\*\*\*\*\*\*\*\*\*\*\*\*\*\*\*\*\*\*\*\*\*\*\*\*\*\*\*

(*) **<u>Authors</u>:**     **Ricardo López-Ruiz**          **(Zaragoza, Spain)**
                 **&**
            **Danièle Fournier-Prunaret**    **(Toulouse, France)**

\*\*\*\*\*\*\*\*\*\*\*\*\*\*\*\*\*\*\*\*\*\*\*\*\*\*\*\*\*\*\*\*\*